\documentclass{PoS}
\usepackage{graphicx}
\usepackage{array}
\usepackage{caption}
\DeclareGraphicsExtensions{.pdf,.png,.jpg}

\title{QUBIC - The Q\&U Bolometric Interferometer for Cosmology - A novel way to look at the polarized Cosmic Microwave Background}

\ShortTitle{QUBIC}

\author{
\speaker{Aniello Mennella}$^{a,b}$,
P.A.R. Ade$^c$,
J. Aumont$^d$,
S. Banfi$^{e,f}$,
P. Battaglia$^g$,
E.S. Battistelli$^{h,i}$,
A. Ba\`u$^{e,f}$,
B. B\'elier$^j$,
D.Bennett$^k$,
L. Berg\'e$^l$,
J.Ph. Bernard$^m$,
M. Bersanelli$^{a,b}$
% ...
\\
\llap{$^a$}University of Milan, Dept. of Physics, Milano, Italy\\
\llap{$^b$}INFN Milano 1 section, Milano, Italy\\
\llap{$^c$}Cardiff University, Cardiff, UK\\
\llap{$^d$}IAS, Orsay, France\\
\llap{$^e$}Universit\`a degli Studi di Milano-Bicocca, Milano, Italy\\
\llap{$^f$}INFN Milano-Bicocca section, Milano, Italy\\
\llap{$^g$}Universit\`a Degli Studi di Trieste, Trieste, Italy\\
\llap{$^h$}Universit\`a di Roma La Sapienza, Roma, Italy\\
\llap{$^i$}INFN Roma 1 section, Roma, Italy\\
\llap{$^j$}IEF, Orsay, France\\
\llap{$^k$}NUIM, Maynooth, Ireland\\
\llap{$^l$}CSNSM, Orsay, France\\
\llap{$^m$}IRAP, Toulouse, France\\
\llap{$^n$}APC, Paris, France\\
\llap{$^o$}Richmond University, Richmond, VA, USA\\
\llap{$^p$}Universit\`a di Roma Tor Vergata, Roma, Italy\\
\llap{$^q$}INFN Roma Tor Vergata section, Roma, Italy\\
\llap{$^r$}University of Manchester, Manchester, UK\\
\llap{$^s$}LAL, Orsay, France\\
\llap{$^t$}ITeDA, CNEA-CONICET-UNSAM, UTN, Argentina\\
\llap{$^u$}Centro Atomico Bariloche, CNEA/CONICET, Argentina\\
\llap{$^v$}Brown University, Providence, RI, USA\\
\llap{$^w$}SISSA, Trieste, Italy\\
\llap{$^x$}University of Wisconsin, Madison, WI, USA\\
\llap{$^y$}IAR, CCT La Plata, CONICET/CIC, Argentina\\
\llap{$^z$}California Institute of Techhnology, Pasadena, 91125 CA, USA\\
\llap{$^{aa}$}Institute of Carlow, Carlow, Ireland
}

\abstract{In this paper we describe QUBIC, an experiment that takes up the challenge posed by the detection of primordial gravitational waves with a novel approach, that combines the sensitivity of state-of-the art bolometric detectors with the systematic effects control typical of interferometers. The so-called ``self-calibration'' is a technique deeply rooted in the interferometric nature of the instrument and allows us to clean the measured data from instrumental effects. The first module of QUBIC is a dual band instrument (150 GHz and 220 GHz) that will be deployed in Argentina during the Fall 2018.}

\FullConference{The European Physical Society Conference on High Energy Physics\\
		5-12 July, 2017\\
		Venice}

\begin{document}

\section{Introduction}
\label{sec_introduction}

	QUBIC is an experiment based on the concept of bolometric interferometry \cite{Battistelli2011} and designed to constrain tightly the $B$-mode polarization anisotropies of the Cosmic Microwave Background (CMB). $B$-mode searches need multi-frequency sensitive instruments to control foreground contamination, and an unprecedented level of control of systematic effects that should be  designed \textit{in-hardware}, as much as possible. 

QUBIC addresses these issues in an innovative way, by combining the advantages of interferometry in terms of systematic effects and the sensitivity of bolometric detectors. The first QUBIC module will operate from the ground observing the sky in two spectral bands centred at 150 and 220\,GHz \cite{qubic15} and will be deployed in Argentina, at the Alto Chorrillos site.

\section{The instrument}
\label{sec_instrument}

	QUBIC is a bolometric interferometer, in which a dual reflector telescope acts as a beam combiner that sums all the signals picked-up by the sky horns onto the detector focal planes. The output of each detector contains interference terms that are the so-called ``visibilities'' of the selected Fourier modes. 

\begin{figure}[h!]
	\begin{center}
    	\includegraphics[width=9cm]{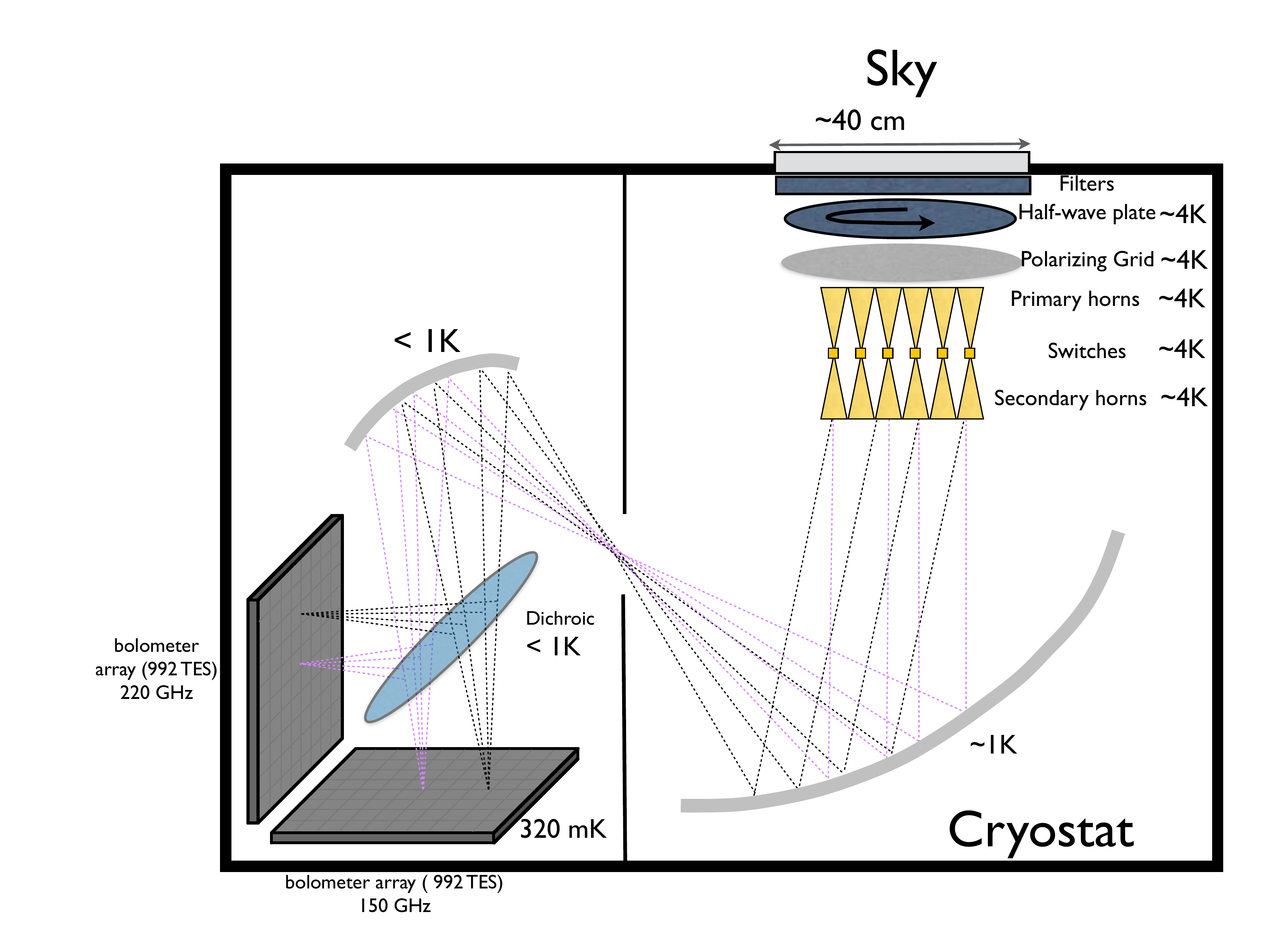}
        \includegraphics[width=6cm]{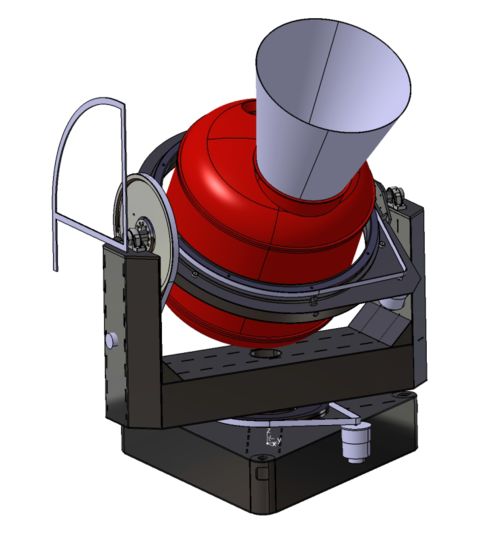}
        \caption{left: schematic of the QUBIC instrument. Right: the QUBIC cryostat integrated on the mount. A forebaffle  is visible on the cryostat aperture window. The mount is approximately 2\,m$\times$2\,m$\times$2\,m in size, the cryostat is about 1.5\,m (height) $\times$ 1.5\,m (diameter) and the forebaffle is about 1\,m high.\label{fig_schematic_qubic}}
    \end{center}
\end{figure}

The schematic in the left panel of Fig.~\ref{fig_schematic_qubic} shows the basic elements in the cryostat. The signal from the sky enters the cryostat through a window and a set of filters. A rotating half-wave plate modulates the  polarization and then one polarization state is selected by a polarizing grid. An array of 400 back-to-back corrugated horns collects the radiation and re-images it on a dual-mirror optical combiner that focuses the signal onto two orthogonal TES detectors focal planes. A dichroic filter placed between the optical combiner and the focal planes selects the two frequency bands, centered at  150\,GHz and 220\,GHz. The right panel of Fig.~\ref{fig_schematic_qubic} shows a 3D rendering of the cryostat integrated on the mount and a fore-baffle mounted around the cryostat window.

A fundamental element is a movable shutter placed in the middle of each back-to-back horn. Each shutter acts as a RF switch (a blade that can slide into a smooth-walled waveguide), that is used to exclude particular baselines when the instrument operates in calibrating mode. We call this particular calibration strategy \textit{self-calibration}, which is a key feature of the QUBIC systematic effects control. The interested reader can find details about self-calibration in \cite{bs13}.
    
\section{Measurement and self-calibration}
\label{sec_measurement_calibration}

	According to the QUBIC design the image that forms on its focal planes is an interference pattern of the radiation reflected by the optical combiner. Therefore, the signal measured at time $t$ by a detector $p$ at the focal plane is:
\begin{equation}
	R(p,\nu,t)=S_I(p,\nu)+\cos(4\,\phi_\mathrm{HWP}(t))\,S_Q(p,\nu)+\sin(4\,\phi_\mathrm{HWP}(t))\,S_U(p,\nu),
    \label{eq_signal_model}
\end{equation}
where $\nu$ is the frequency and $\phi_\mathrm{HWP}$ is the angle of the half-wave plate at time $t$. The three terms $S_{I,Q,U}$ in Eq.~\ref{eq_signal_model} represent the sky signal in intensity and polarization convolved with the so-called \textit{synthetic beam}.

\begin{figure}[h!]
  \begin{center}
      \includegraphics[width=16cm]{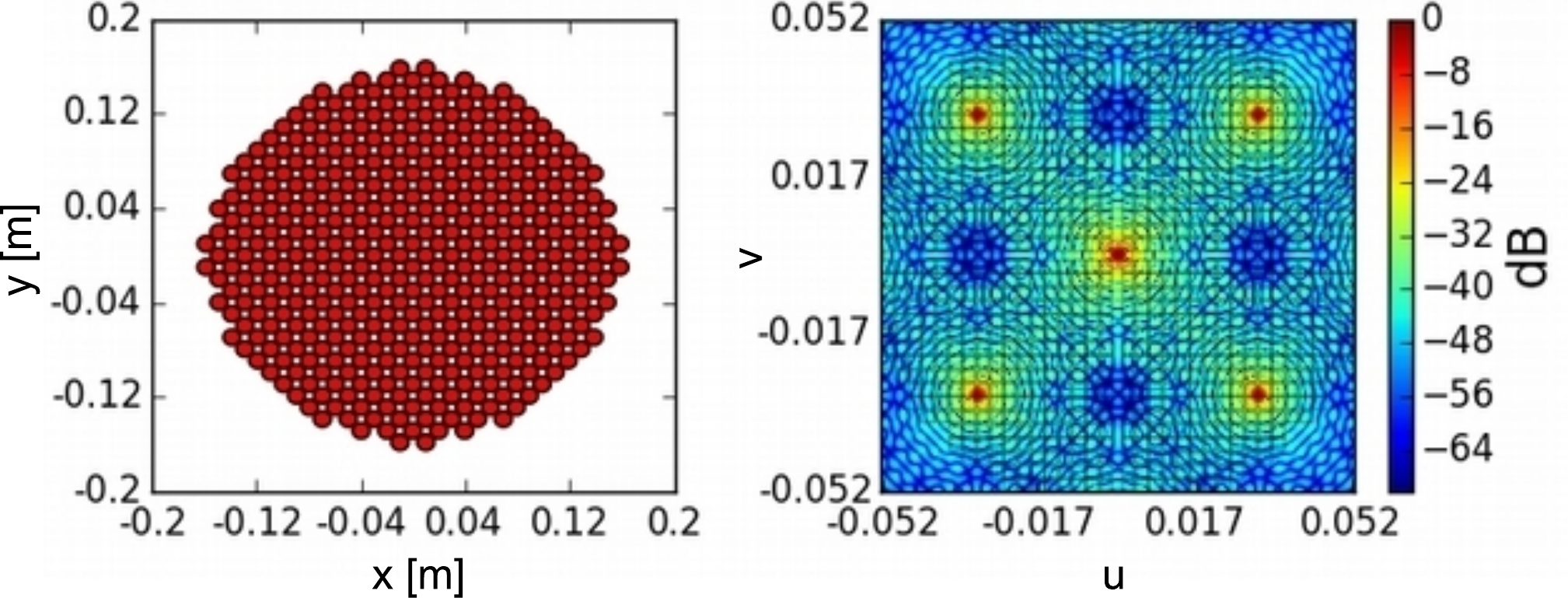}
      \caption{\label{fig_synthetic_beam}left: QUBIC aperture plane showing all 400 antennas open to the sky. Right: the interference pattern formed on each of the focal planes when the instrument is observing a far point source located vertically along the instrument line-of-sight. The $u$ and $v$ coordinates are defined as: $u=\sin\theta\cos\phi$ and $v=\sin\theta\sin\phi$, where $\theta$ and $\phi$ are the angles on the celestial sphere defining the synthetic beam.}
  \end{center} 
\end{figure}

The images in Fig.~\ref{fig_synthetic_beam} help the reader to understand qualitatively this concept. Imagine that QUBIC observes a point source in the far field located directly along the line-of-sight of the instrument with all the 400 antennas open to the sky. The image formed on each of the focal planes (see the right panel of Fig.~\ref{fig_synthetic_beam}) is an interference pattern formed by peaks and lobes. This pattern actually works like a \textit{beam pattern} that convolves the sky signal.

Therefore, if $X$ is the signal from the sky (either in intensity, $I$, or polarization, $Q,U$), then the measured signal on the pixel $p$ is the convolution between the signal and the synthetic beam pattern of pixel $p$. In mathematical form: $S_X(p) = \int X(\mathbf{n})B^p_\mathrm{synth}(\mathbf{n})d\mathbf{n}$. This means that QUBIC data can be analyzed exactly like the data obtained from a normal imager, provided that we build a window function of the synthetic pattern for each pixel. 

A feature that is unique in QUBIC compared to normal imagers is the systematic error control provided by the possibility to calibrate independently each interferometric baseline. This feature is named \textit{self-calibration} and it is based on the possibility to open and close any of the 400 instrument apertures.
In the self-calibration mode pairs of horns are successively shut while QUBIC observes an artificial partially polarized source. Then we reconstruct the signal measured by each individual pairs of horns in the array and compare them. The point now is that redundant baselines correspond to the same mode of the observed field, so that a different signal between them can only be due to photon noise or instrumental systematic effects. Using a detailed parametric model of the instrument we can fully recover the instrument parameters through a non-linear inversion process. The updated model of the instrument can then be used to reconstruct the synthetic beam and improve the map-making, reducing the leakage between Stokes parameters (see Fig.~\ref{fig_self_calibration}).

\noindent
\begin{minipage}{.5\textwidth} 
Figure ~\ref{fig_self_calibration}: improvement in the power spectrum estimation exploiting self calibration according to three calibration schemes. Even with 1\,s per baseline (corresponding to a full day dedicated to self-calibration) we can reduce significantly the $E\rightarrow B$ leakage. This leakage can be further reduced by spending more time in self-calibration (adapted from \cite{bs13}). The parameter $r$ represents the ratio between the amplitudes of the tensor and scalar fluctuations during inflation.
\end{minipage}
\begin{minipage}{.5\textwidth}
\begin{center}
		\includegraphics[width=6cm]{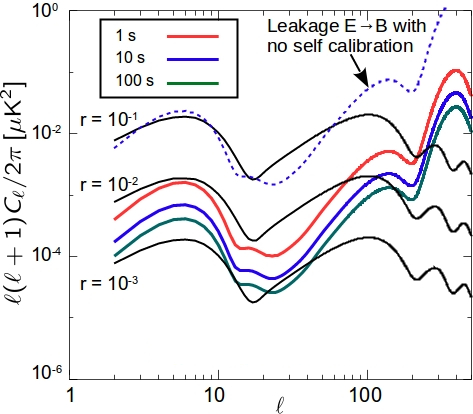}
		\label{fig_self_calibration}
		\end{center}    
\end{minipage}

%

\section{The QUBIC site and the expected scientific performance}
\label{sec_qubic_site}

	The QUBIC first module will be installed in Argentina, near the city of San Antonio de los Cobres, in the Salta Province. This site has coordinates (24$^\circ$ 11$’$ 11.7$''$ S; 66$^\circ$ 28$’$ 40.8$''$W) and an altitude of 4869\,m a.s.l.

The left panel in Fig.~\ref{fig_site_quality} (adapted from \cite{Aumont2016}) shows the overall site quality. In the plot we see the uncertainty in the tensor-to-scalar ratio, $r$, as a function of the fraction of usable time in the two sites. The circled point shows the estimate for the case in which 12 hours a day are spent in calibration mode. The right panel in Fig.~\ref{fig_site_quality} summarizes the main scientific performance of QUBIC in terms of tensor-to-scalar ratio final uncertainty. In particular the figure shows that with two years of continuous observations with the first module it is possible to reach a value for $\sigma_r$ of the order of 10$^{-2}$.
\setcounter{figure}{3}
\begin{figure}[h!]
	\begin{center}
		\includegraphics[width=7cm]{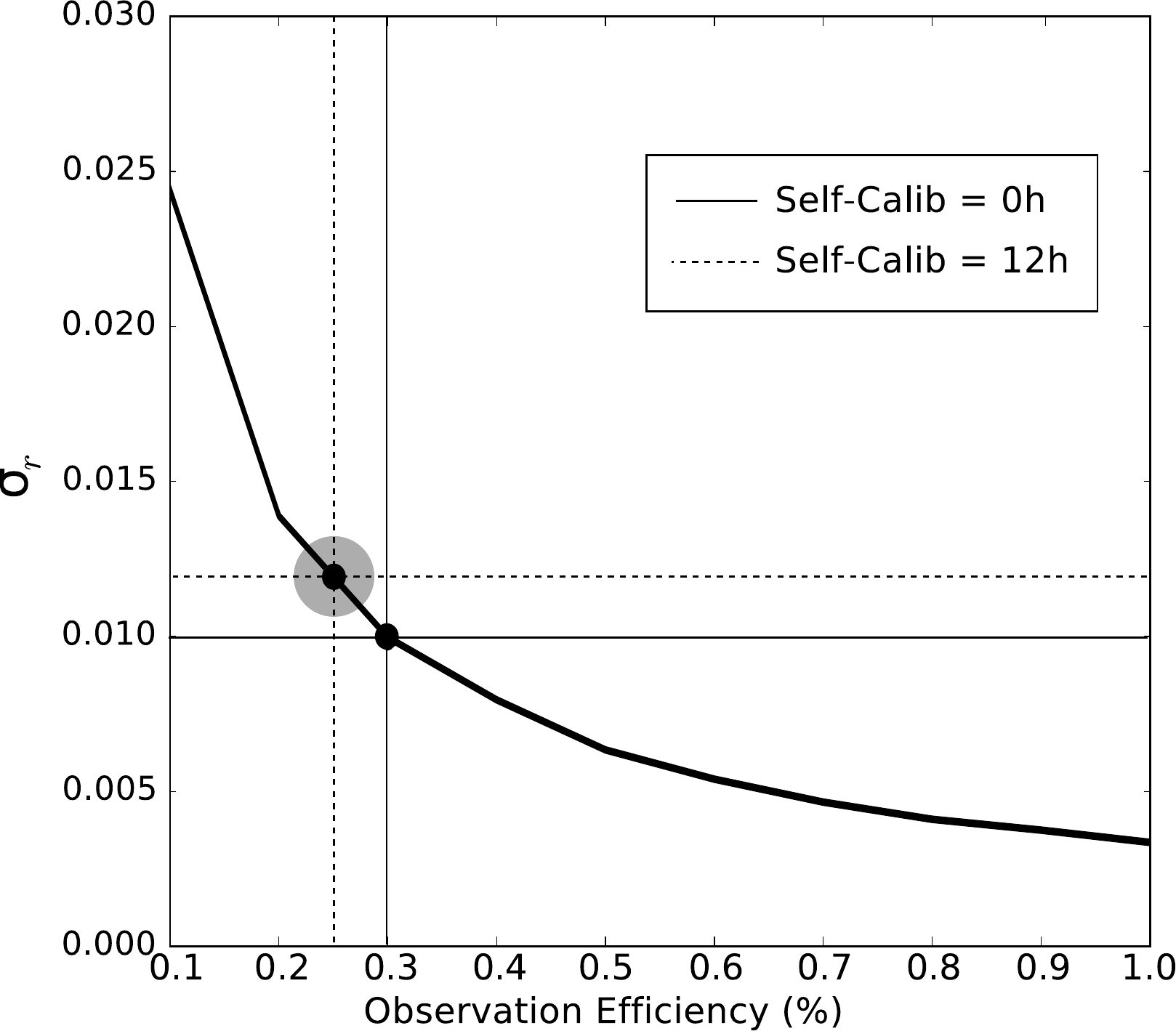}
        \includegraphics[width=7.4cm]{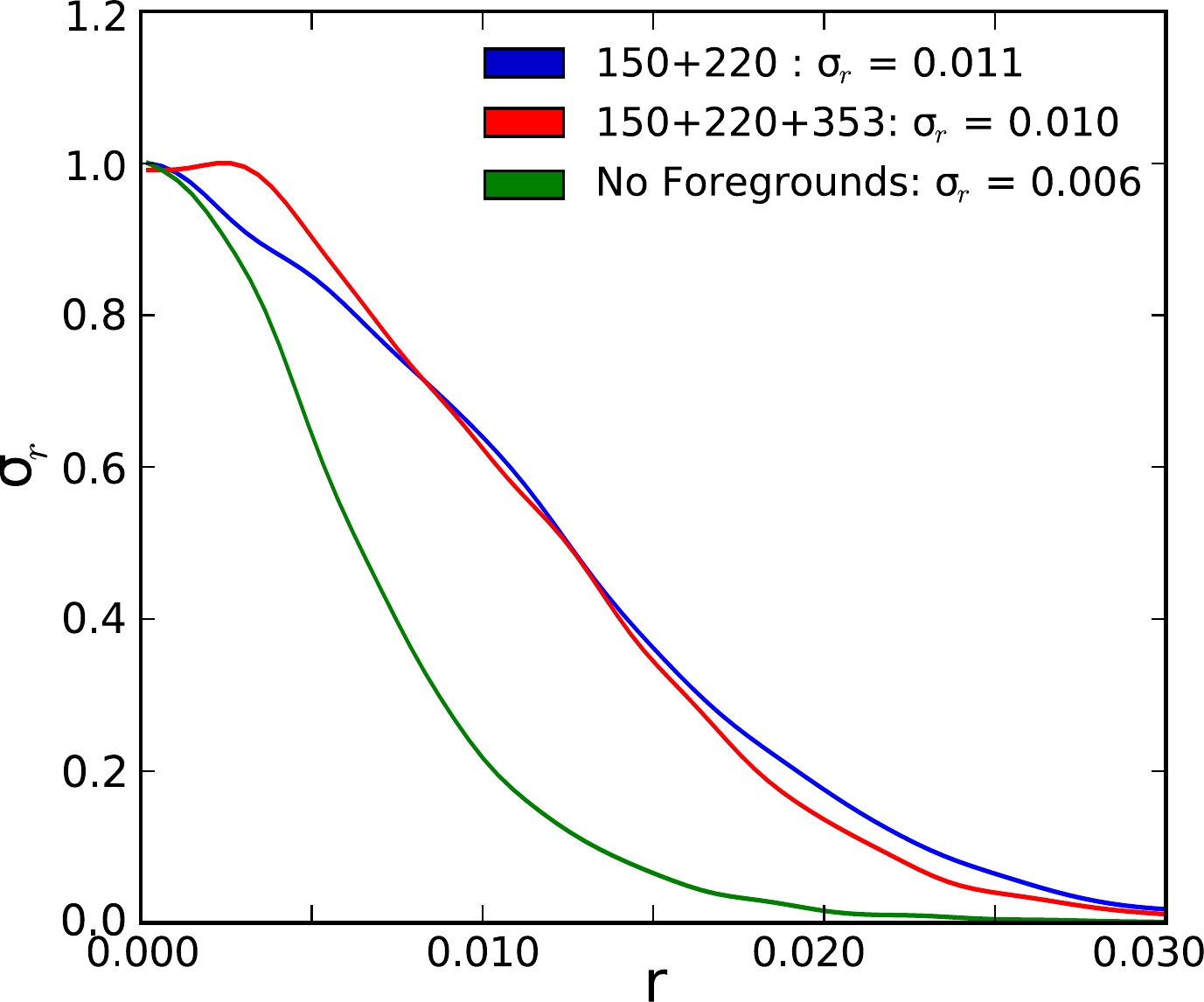}
        \caption{left: assessment of the site quality in Alto Chorrillos for QUBIC in terms of $\sigma(r)$. The circle corresponds to the case in which 12 hours a day are spent in self-calibration mode. Right: expected forecasts on the tensor-to-scalar ratio with QUBIC using a full likelihood with three configurations: QUBIC alone (blue), QUBIC with Planck 353 GHz information added (red), QUBIC if no foregrounds were present (green). }
        \label{fig_site_quality}
    \end{center}
\end{figure}
    

    
\section{Conclusions}
\label{sec_conclusions}

	QUBIC represents a new concept of measuring the polarization of the CMB. It combines the high sensitivity of TES bolometric arrays with the control of systematic effects that is typical of interferometers. This is a key asset in this kind of measurements, where the quest for high sensitivity must be backed by instruments that are inherently immune to spurious effects. 
QUBIC responds to this challenge with the key feature of \textit{self-calibration}, that is possible thanks to the interferometric nature of the instrument. The price to pay is an instrument that is technologically more challenging compared to an imager and a more complex optical response (the \textit{synthetic beam} that must be used to deconvolve the raw data).
The first QUBIC model that will be deployed in Argentina during 2018 will demonstrate the feasibility and potential of this approach and open the way for a new generation of instruments in the field of Cosmic Microwave Background polarimetry.

\bibliographystyle{JHEP}
\bibliography{biblio}

\end{document}